\documentclass[
  english,        
  font=times,     
  onecolumn,      
]{tumarticle}

\usepackage{placeins}
\usepackage{amsmath}
\usepackage{csquotes}
\usepackage{float}
\usepackage{booktabs}
\usepackage{multirow}
\usepackage{multicol}
\usepackage{siunitx}
\usepackage{subcaption}
\usepackage{colortbl} 
\usepackage{chemformula} 

\usepackage{xr} 
\makeatletter
\newcommand*{\addFileDependency}[1]{%
  \typeout{(#1)}%
  \@addtofilelist{#1}%
  \IfFileExists{#1}{}{\typeout{No file #1.}}%
}
\makeatother

\newcommand*{\myexternaldocument}[2][]{%
  \externaldocument[#1]{#2}%
  \addFileDependency{#2.tex}%
  \addFileDependency{#2.aux}%
}
\myexternaldocument{SI}

\usepackage[noabbrev]{cleveref}       

\DeclareSIUnit\angstrom{\text {Å}}
\DeclareSIUnit\rydberg{\text{Ry}}
\DeclareSIUnit\hartree{\text{Ha}}
\DeclareSIUnit\bohr{\text {\ensuremath {a}}_{0}} 

\usepackage{comment}
\usepackage{bm}
\usepackage{chemformula}
\usepackage[natbib=true,
    style=numeric,
    sorting=none
]{biblatex}
\usepackage{geometry}
\usepackage{tabularx}
\usepackage{array} 
\usepackage{caption} 
\usepackage{makecell} 

\title{AUGUR, A flexible and efficient optimization algorithm for identification of optimal adsorption sites}

\author[affil=1, email=ioannis.kouroudis@tum.de]{Ioannis Kouroudis $^*$}
\author[affil=1]{Poonam \footnote{Joint first author}}
\author[affil=1]{Neel Misciaci}
\author[affil=1]{Felix Mayr}
\author[affil=1]{Leon M{\"u}ller}
\author[affil=1]{Zhaosu Gu}
\author[affil=1, email=alessio.gagliardi@tum.de]{Alessio Gagliardi}

\affil[mark=1]{Chair of Simulation of Nanosystems for Energy Conversion, Department of Electrical Engineering, TUM School of Computation, Information and Technology, Atomistic Modeling Center (AMC), Munich Data Science Institute (MDSI), Technical University of Munich, Hans-Piloty-Straße 1, 85748 Garching, Germany}

\bibliography{ref}

\begin{document}

\maketitle

\begin{abstract}

  In this paper, we propose a novel flexible optimization pipeline for determining the optimal adsorption sites, named AUGUR (Aware of Uncertainty Graph Unit Regression). Our model combines graph neural networks and Gaussian processes to create a flexible, efficient, symmetry-aware, translation, and rotation-invariant predictor with inbuilt uncertainty quantification. This predictor is then used as a surrogate for a data-efficient Bayesian Optimization scheme to determine the optimal adsorption positions. This pipeline determines the optimal position of large and complicated clusters with far fewer iterations than current state-of-the-art approaches. Further, it does not rely on hand-crafted features and can be seamlessly employed on any molecule without any alterations. Additionally, the pooling properties of graphs allow for the processing of molecules of different sizes by the same model. This allows the energy prediction of computationally demanding systems by a model trained on comparatively smaller and less expensive ones.
\end{abstract}

\section{Introduction}

Novel, functional structures at the nanoscale could be crucial for transforming a broad spectrum of economically significant processes into greener and more sustainable solutions. For instance, nanostructured materials hold the potential to significantly enhance the cost-effectiveness of fuel-cell devices \cite{kouroudis2023utilizing}, enable the creation of highly efficient quantum-dot LEDs \cite{lampe2023rapid}, and pave the way for generating atom-precise efficient nanocatalysts for studying novel catalytic pathways in electrochemical applications \cite{Mechanistic_Poonam, Marlon}.

As performance is highly dependent on specific structural characteristics which often can not easily be resolved in lab experiments, computational chemistry - most often by using Density Functional Theory (DFT) based approaches - can be used to generate in-silico insights.
Typical questions range from elucidating which feature of a given nanoparticle might improve catalytic performance to mechanistic explanations for key synthesis procedures, allowing tailored experiments to drive up experimental yields for optimal structures.

Commonly, these questions are associated with finding energetically favorable configurations for the potential energy surface (PES) of a system, which is a property relevant to solving a wide range of problems in computational chemistry.

The established methodology allows finding "docking" mechanisms between small molecules and large biomolecules, which is relevant for drug development \cite{morris2009autodock4}. Additionally, a large area of research revolves around the sensing of harmful gases by novel nanomaterials chosen according to their strength of interactions. In the field of catalysis, identifying and maximizing relevant catalytic sites with the lowest number of atoms results in highly efficient and financially viable catalysts \cite{Sklason2007}. 

For the latter, the primary approach starts by finding symmetry points of a surface crystal plane and sampling a small area of the configuration space around them \cite{Wei1998a} to find a minimal energy configuration. 
All system constituents are usually kept rigid in this approach, and only the final configurational pick is subject to a relaxation procedure and further study. 
Over time, several physics-based optimization techniques have evolved to allow for a more elaborate study of the PES moving beyond simple sampling of enumerated configurations: 
the Nudged Elastic Band (NEB) method allows the description of catalysis pathways \cite{jonsson1998nudged}.  With Minima-Hopping and genetic algorithms, new stable material conformations can be explored \cite{Goedecker2004, Schnborn2009}, and metadynamics could be used to examine the thermodynamics of a given system \cite{barducci2008well, YazdanYar2018}. 
Despite these methodological improvements, the high number of necessary energy evaluations has relegated those to niche and specialist applications. At the same time, the original sampling-based workflows for configurational search are still in everyday use and can be performed in a highly automated fashion \cite{Marti2021, Pedretti2023}. 
However, when applied to novel systems incorporating irregularly shaped nanostructures or more complex adsorbents, exhaustive sampling quickly reaches the computational limits of the average project. 
Here, data-driven machine learning techniques such as neural networks have been highly utilized \cite{mayr2022machine} and could be used to describe the PES and subsequently allow for more efficient exploration. 
While early research showed that this approach could successfully reproduce even complex PES \cite{Lorenz2006, jager2018machine}, the practical application requires that this can be achieved without exhaustively sampling the PES of the system under examination for data generation.

One approach is using a large, pre-trained model describing the system and using that to significantly speed up an exhaustive sampling and relaxation procedure, as has been done to model adsorption dynamics \cite{PabloGarca2023, Lan2023}. 
However, this relies on the availability of a large-scale dataset to create a model with suitable generalization power towards the research problem. 
Without such prerequisites, one can also iteratively build up an uncertainty-quantifying model for the PES of a specific system using techniques such as active learning (AL) or Bayesian optimization (BO). 
The "classic" approach to this problem would be a Gaussian-process (GP) based regression model for the PES. The inputs would be features extracted from the atomistic structure \cite{Jinnouchi2019, Xie2023}, with newer research suggesting the usage of end-to-end-trained graph-neural-network models \cite{wollschlager2023uncertainty}. 
These kinds of machine-learning models have been used to find stable crystal structures \cite{Yamashita2018, Deshwal2021a, Zuo2021}, learn adsorbate-surface PES \cite{Tran2018a} and have been integrated into techniques like NEB \cite{Peterson2016, Koistinen2019} or minima-hopping \cite{Jung2023} resulting in significant computational speedup.

The basic problem of finding energetically minimal configurations for a variety of systems using BO has been extensively studied under the \texttt{BOSS}-handle using problem-specific features and Gaussian process-based PES surrogates. Starting by demonstrating that energetical minimal molecular structures can be found by optimizing a model based on dihedral angles \cite{Chan2019}, the method was further demonstrated for adsorption problems by finding the optimal placement of large, rigid molecules on inorganic surfaces \cite{todorovic2019bayesian, Jaervi2020} and has been refined to determine optimal adsorption structures for partly-flexible molecules on gold-clusters \cite{Fang2023}. 
They, however, rely on hand-crafted feature extraction functions, which can restrict the potential expressibility of the model and require a high level of prior physical understanding that might not be present in more complex systems. Further, this approach isn't transferable as it relies on a fixed, molecule-dependent size of input features. 

Our method encodes the cluster-adsorbate system as a graph. This is processed by a state-of-the-art Graph Neural Network (GNN), which allows the representation to be symmetry, rotation, and translation invariant. Given the pooling properties of the graph, this output has the same dimensionality independently of the molecule used for input. The output of the GNN is used as input to a Gaussian Processes (GP) model which creates the final predictions along with quantification of their uncertainty. Finally, a Bayesian Optimization (BO) scheme produces suggestions for optimal adsorption sites.

We demonstrate the robustness of the AUGUR framework on two systems. The first is the family of the chini clusters that have been exhaustively studied as a precursor to generating atom-precise nanoclusters using Molecular Organic Framework as a template in our previous study \cite{Mechanistic_Poonam}. The atom-precise nanoclusters are crucial in developing the next generation of highly tailored, efficient, and economically viable nanocatalysts. The DFT simulations of Zn$^{2+}$ - Chini cluster interactions found in literature allow us to know, a priori, the energetic behavior of all chemically distinct sites. This, in turn, allows the validation of our algorithm.  
Second, we selected ZnO clusters that are well-known experimentally to be used in semiconductor companies for chips, sensors, and electrodes due to their observed high catalytic activity and stability \cite{Zno_applications}. In our study, the focus lies on the gas adsorption capability of the \ch{(ZnO)78} cluster, as it is one of the "magic clusters"\cite{ZnO78_magic_cluster}. However, due to the large size of the cluster, the conventional approach of exhaustive DFT investigation, covering the entire surface of the cluster, is computationally infeasible. 
The AUGUR pipeline has proven itself robust, transferable, and highly efficient in both case studies, identifying the optimal sites with approximately ten simulation runs and without the need for hand-crafted features.

\section{AUGUR pipeline}
\begin{figure}[h!]
    \centering
    \includegraphics[width = 0.9\textwidth]{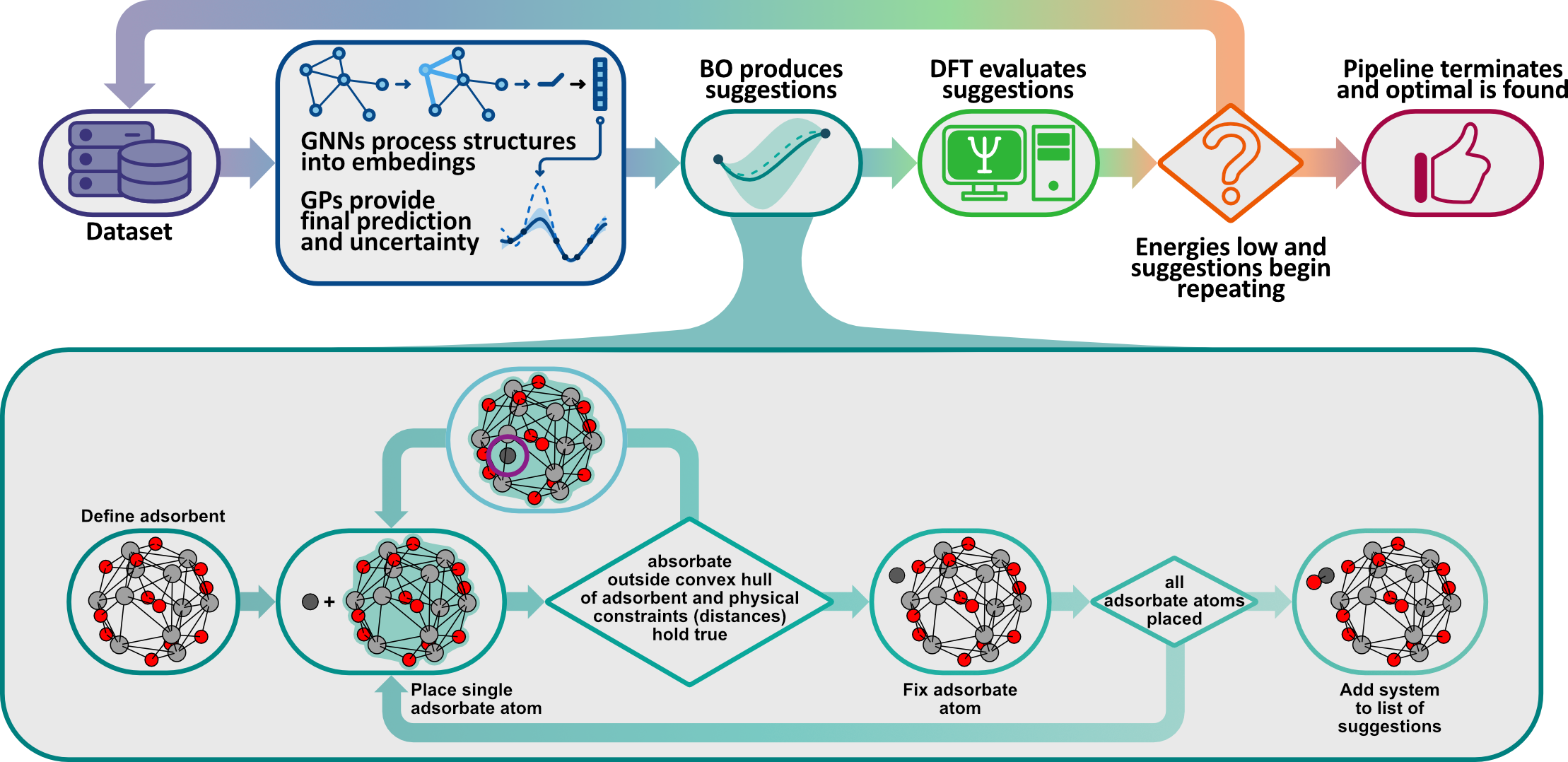}
    \caption{AUGUR pipeline summary. The top line is the optimization pipeline, from left to right, training the GNNs and the GPs, using them as surrogates for BO, evaluating the BO suggestions with DFT, adding the new results into the dataset, and repeating them. The bottom line is the point generation for BO, from left to right, define the cluster, place the first adsorption atom at a physically consistent distance and outside the convex hull of the molecule, and repeat this process with all atoms of the adsorbant molecule.}
    \label{fig:enter-label}
\end{figure}

Our pipeline consists of four major components, also visualized in \Cref{fig:enter-label}:
\begin{itemize}
    \item A graph neural network that is responsible for extracting a meaningful, rotation, translation, and symmetry invariant representation of the molecule. The output dimension of the graph is fixed regardless of the molecule's size and, therefore, can be used across various molecules.
    \item A Gaussian Process that receives as input the output of the graph and predicts the interaction energy of the system and the corresponding uncertainty of the prediction.
    \item A Bayesian Optimizer that uses the above two models as a combined surrogate and generates suggestions for adsorption positions. These are simulated by the simulation model.
    \item A Density Functional Theory model that provides a physically accurate but computationally expensive evaluation of the interaction energy of the suggestions. 
\end{itemize}
\FloatBarrier
\subsection{Density Functional Theory}
Density Functional Theory (DFT) is an approximation method used to solve the Schrödinger equation for many-body systems, based on the hypothesis proposed by P. Hohenberg and W. Kohn \cite{Hohenberg_kohn_theorem}. The Hohenberg-Kohn (HK) hypothesis asserts that the ground state observables of the Schrödinger equation are functionals of the electronic density of the system rather than the $\mathrm{3N}$ coordinates of the particles. 





The total energy (functional of density) of the system is given by:

\begin{equation}
E[\rho] = T_{\mathrm{s}}[\rho] + E_{\mathrm{H}}[\rho] + E_{\mathrm{ext}}[\rho] + E_{\mathrm{xc}}[\rho]
\end{equation}

\noindent where $T_{\mathrm{s}}[\rho]$ is the kinetic energy of the non-interacting electrons, $E_{\mathrm{H}}[\rho]$ is the Hartree energy representing the electron-electron repulsion, $E_{\mathrm{ext}}[\rho]$ is the external potential energy from the nuclei, and $E_{\mathrm{xc}}[\rho]$ is the exchange-correlation energy that includes all complex quantum mechanical effects.

The variational principle used to determine the ground-state energy is given by:

\begin{equation}
E_0 = \min_\rho \left\{ E[\rho] \right\}
\end{equation}

Density Functional Theory (DFT) is particularly valued for its computational efficiency. Unlike traditional ab-initio methods such as Hartree-Fock (HF), which are dominated by the computation of two-electron integrals that scale with the fourth power of the system size ($N^4$), DFT scales more favorably at $N^3$. In this context, N represents the system size, encompassing the number of atoms, electrons, or basis functions. Consequently, if the system size doubles, the computational effort for HF increases by a factor of 16, whereas for DFT, it only increases by a factor of 8. This more favorable scaling allows DFT to perform calculations faster and handle larger systems more effectively than traditional ab-initio methods.

Furthermore, DFT inherently accounts for electron correlation effects, which are often neglected by HF. This capability makes DFT especially useful for systems where electron correlation is significant, such as in transition metal chemistry, which is central to many chemical reactions and the use cases in our study.

Despite the advantages, DFT does have limitations, particularly in terms of system size. Although DFT simplifies the many-body problem of N electrons to a dependency on the electronic density (reducing the complexity to three spatial coordinates), it is generally constrained to handling systems with a few hundred atoms. The computational effort still scales approximately as $N^3$, making the study of electronic structure properties of large systems resource-intensive and time-consuming.

To mitigate these limitations, our study leverages AUGUR, a framework that guides the experimental process using Bayesian Optimization (BO) enhanced by graph-based Gaussian Processes. This approach reduces the number of required DFT simulations, significantly lowering the computational burden.

\subsection{Bayesian Optimization}
Bayesian Optimization (BO) is an optimization algorithm that has proven itself to be both data efficient and accurate even in non-convex/concave optimization problems. 
In principle, let us consider a costly optimization problem, such as one relying on expensive simulations or arduous experimental work for the evaluation of the objective function. The natural solution to cases such as these is the training of a surrogate model that will provide sufficient accuracy at a fraction of the required time. Nevertheless, data-driven models typically require a lot of data to be trained sufficiently, which in turn invalidates the main motivation of the surrogate model, i.e., the minimization of the requirement of the time-consuming part of the pipeline. Bayesian optimization is particularly suited to this kind of problem because it simultaneously trains a surrogate model efficiently and identifies hopeful query points. 
It relies on two components. The first is a stochastic predictor, i.e., a predictor that can provide uncertainty quantification, typically Gaussian Processes (GP), Bayesian Neural Networks (BNN), or others. The second is a function that evaluates each possible point on a) how optimal it is and b) how much new information it injects into the model. This function shall be referred to as the acquisition function. For the remainder of the chapter, we will describe a minimization process, but everything can be applied without loss of generality to any optimization problem. 
In the present work, we seek to find the position of adsorption that will result in the lowest possible energy. To this end, we can generate a number of random positions and evaluate their respective energies using the uncertainty-aware predictor. The energy value is an indicator of the site's optimality. The standard deviation reflects how uncertain the prediction is. Choosing to focus on the points of high standard deviation will lead to the most efficient injection of new information and, therefore, the fastest training of the surrogate model. This is called exploration and is typically the focus of the first queries of the ground truth process. As the surrogate improves, the focus shifts to determining the optimal points, and therefore, the choice is increasingly being determined by the prediction and less by the standard deviation. This phase is called exploitation. 
A very intuitive example of an acquisition function is the lower confidence bound (LCB):
\begin{equation}\label{eq:ucb}
  \mathbf{x_{opt}} =  \arg \min_{\mathbf{x}}( \mu(\mathbf{x}) - \epsilon \sigma(\mathbf{x}))
\end{equation}
where $\mathbf{\mu}$ and $\mathbf{\sigma}$ are the predictions for the mean and standard deviation, respectively, of the predictor on the corresponding point $\mathbf{x_{opt}}$ and $\epsilon$ is a trade-off parameter. The parameter $\epsilon$ is directly linked to the \textit{exploration} and \textit{exploitation} phase of the optimization process.
 In this work, the so-called expected improvement acquisition function was utilized:
\begin{equation}\label{eq:expected_improvement}
  EI(x)= \mathbb{E}\left [\max\left( f(\mathbf{x}^+) - f(\mathbf{x}),\, 0 \right) \right]
\end{equation}
where the superscript $+$ denotes the best point so far and the function $f$ is the trained surrogate model based on Gaussian processes. A complete derivation of \Cref{eq:expected_improvement} can be found in Ref. \cite{brochu2010} along with resources on additional acquisition functions.
For completeness, we note that other acquisition functions, such as probability of improvement and LCB, were also implemented but yielded inferior results compared to expected improvement.\\

Note that there are multiple, much more efficient ways of optimizing the acquisition functions, including gradient-based and evolutionary algorithms. Nevertheless, the complexity of the constraints of this problem made their application non-trivial. In comparison, random point generation proved very efficient, especially given that the deep kernel method allows for the fast and parallel processing of thousands of points in mere seconds. More information about how the points were generated can be found in \Cref{chap:point_generation}.

\subsection{Point generation}
The convergence of Bayesian Optimization can be significantly sped up if we introduce a layer of physical understanding of the problem. The simplest way is to generate a number of points (in our case, 10.000) that adhere to a set of constraints. 
A random atom of the cluster is chosen. The adsorbing atom is placed on the surface of a sphere with a radius determined by the physical limits of the interaction distance between the two atom types. Then, the adsorbate is tested for being placed inside the cluster or outside. This is done by performing a Delaunay tessellation on the cluster coordinates and determining the location of the adsorbent with respect to the resultant convex hull. If the adsorbent is found inside the hull, it is discarded, and the process starts anew. If this constraint is not violated, the adsorbent distance to the remaining cluster atoms is determined. If it was not positioned closer than the acceptable limits for any atom, then the new adsorbent position is accepted into the sample set and evaluated through the acquisition function. Otherwise, it is discarded, and the process starts anew. The sample generation process can be summarized as follows:

\begin{enumerate}
    \item Choose a random atom of the cluster, provided adsorption is physically possible on it.
    \item Place the adsorbent on the surface of a sphere with a radius determined by physical knowledge of the maximum and minimum interaction distance of these two atoms.
    \item Test if the adsorbent falls inside the convex hull of the cluster. 
    \item If the above constraint is not violated, test if the adsorbent is closer than a distance of tolerance to the other atoms.
    \item If the above constraint is not violated, the sample is admitted into the list of viable samples.
\end{enumerate}  
If the adsorbent consists of more than one atom, we place every subsequent atom on a sphere with a radius chosen based on the expected bond length and repeat steps 3 and 4 until all the adsorbent atoms have been placed. Then, the total system is accepted into the list of viable samples. A concise flowchart of the process can be found in \Cref{fig:enter-label}.

\FloatBarrier \label{chap:point_generation}
\subsection{Gaussian Processes}
In the current project, the surrogate model chosen was Gaussian processes as it combines robustness to overfitting and the well-documented accuracy of global kernel methods. It is a data-driven stochastic algorithm that models predictions as the posterior of the Bayes formula. The prior and the likelihood are modeled as Gaussian distributions whose parameters are optimized based on the already measured samples. In this way, the algorithm elegantly provides an inherent uncertainty quantification, which in data sets of small size can be crucial. 
The fundamental principle of the method is that the true underlying physical process $\mathbf{y}$ can be modeled by a model $\mathbf{f}$, tuned on a set of points $\mathbf{X}$ using a multivariate Gaussian distribution. This is the prior distribution, and without loss of generality, we can assume it is centered around zero. This assumption can be made more robust by standardizing the data so that their targets also fall within a zero mean Gaussian distribution. The covariance matrix can be defined by the use of a kernel function. A kernel is a distance-like user-chosen function that encodes the correlation of two outputs given the distance of their respective inputs. The prior would then take the mathematical form shown in
\begin{equation}\label{eq:prior}
  \mathbf{f}|\mathbf{X} \sim \mathcal{N}(0,\mathbf{K_{x,x}})
\end{equation}
where $\mathbf{K_{x,x}}$ is the kernel function applied pairwise to all input pairs to construct the covariance matrix.

The likelihood of the observations given the model can be represented as a noisy normal distribution around the model predictions:
\begin{equation}\label{eq:likelihood}
  \mathbf{y}|\mathbf{f} \sim \mathcal{N}(\mathbf{f},\mathbf{\sigma^2I})
\end{equation}
\Cref{eq:prior} and \Cref{eq:likelihood} can be combined through the Bayes formula (\Cref{eq:bayes}) to compute the posterior distribution which will serve as our prediction.
\begin{equation}\label{eq:bayes}
  P(\theta|\textbf{D}) =  \frac{P(\textbf{D} |\theta)P(\theta )}{P(\textbf{D})}.
\end{equation}
Solving the above process analytically parametrizes the posterior prediction and generates the prediction distribution.
\begin{subequations}
  \begin{align}
    \mathbf{\mu_{x^\star}}      & = \mathbf{K_{f^\star,f}[K_{f,f}+\sigma^2I]}^{-1}\mathbf{y} \label{eq:posterior_m}                                            \\
    \mathbf{\sigma_{x^\star}^2} & = \mathbf{K_{x^\star,x^\star}} - \mathbf{K_{x^\star,x}[K_{f,f}+\sigma^2I]}^{-1}\mathbf{K_{x,x^\star}} \label{eq:posterior_s}
  \end{align}
\end{subequations}
where the points for prediction and training are denoted with and without the superscript $\star$, respectively. 
Gaussian processes only require the definition of the prior mean and covariance matrix. The first can be routinely assumed to be zero, though other variations exist, such as a constant mean or a mean that linearly depends on the input values. The latter can be defined with a kernel function whose parameters should, in principle, be uncorrelated to the observations. However, it is common practice to optimize these parameters to maximize the marginal log likelihood to achieve better prediction results. 
Note that a complete mathematical analysis of Gaussian processes and their derivation is beyond the scope of this work. A comprehensive description can be found in Ref. \cite{williams2006}.
\subsection{Graph Neural Networks}
Gaussian processes are a versatile family of algorithms. Nevertheless, they suffer significantly with feature spaces of large dimensionality. Further, meaningfully representing a molecular structure using a set of scalar features is non-trivial. One data-driven algorithm, however, that provides a natural encoding for molecules is Graph Neural Networks (GNNs). 

The inputs of this architecture are graph structures, which consist of nodes, edges, and edge attributes. Each node is a mathematical representation of an atom. Node features that are used in this work are the atomic radius, atomic mass, and electronegativity of each atom. Edges are tuples that describe which nodes are connected. Lastly, the edge attributes contain information about the strength of interaction between two nodes. In the present study, the relevant entry of the coulomb matrix and the pairwise distance between two atoms were used as edge features. 

The resulting representations are processed with graph convolution layers. These layers operate in two steps. 
Initially, all the nodes connected to a specific node create messages that codify the effect of these nodes on the receiving one. The messages can be generated by a variety of algorithms, but in the present work, we chose a neural network inspired by \cite{NNCONV_2}. This process is repeated for every node. 

The second step is the aggregation process, where all the messages directed at one node are condensed into one update that is then applied to the node features. Typically, this can be done by averaging or selecting the maximum message values, but more involved strategies like transformers can be applied. The updated node features now contain information about their neighboring nodes, thus giving more context to the representation.

Finally, once sufficient context has been applied, the graph representation is pooled down to a fixed dimension vector. Typically, this can be done by averaging or selecting the maximum value of every feature across the entire graph. This step condenses a graph of arbitrary size down into a fixed number of representations. In this way, the same algorithm can process different molecules of various sizes without relying on suboptimal techniques.

Graph neural networks can naturally encode molecules but lose the stochastic nature of the Gaussian processes. To this end, we have used a GNN as a feature extractor from the molecules, and these features are used as inputs to the GP. The parameters of the full pipeline are trained simultaneously towards the goal of minimizing the marginal log likelihood. In this way, we combine the advantages of both algorithms.
\section{Results}

\subsection{Chini Clusters}
\begin{figure}[ht]
    \includegraphics[width=\linewidth]{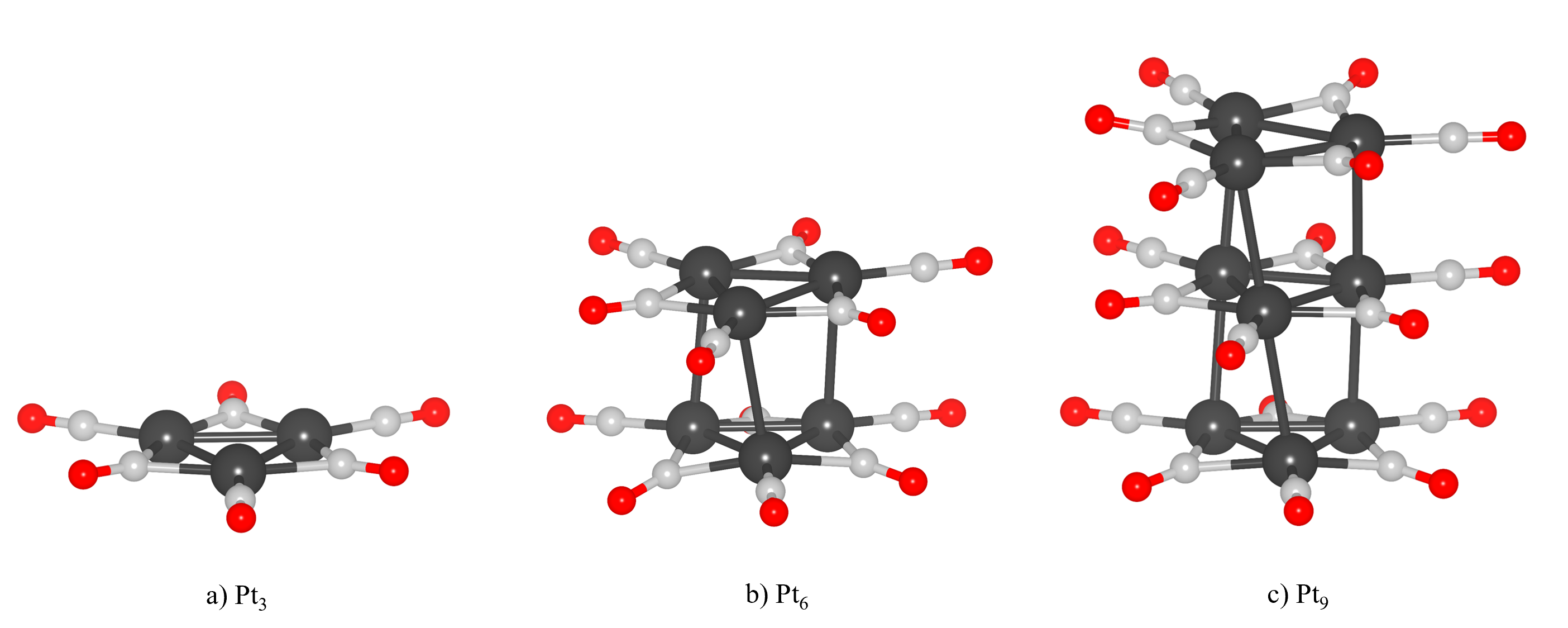}%

  \caption{Case Study 1: Chini clusters $\mathrm{[Pt_{3n}(CO)_{6n}]^{2-}}$ (n = 1–3); These nanoclusters comprise of three Pt-Pt bonds forming a triangle and laterally protected with CO ligands. As n increases, layers are progressively added, as seen in a), b), and c). The color scheme used is as follows: Pt (black); O (red); C (light grey). }
      \label{fig:pt_clusters}

\end{figure}

Chini clusters are a unique class of organometallic compounds known for their unusual bonding and electronic properties. These clusters serve as precursors for synthesizing atomically precise nanoclusters, with their model structures illustrated in \Cref{fig:pt_clusters}. Understanding the interaction between Chini clusters and \ch{Zn^{2+}} ions is crucial for unraveling the encapsulation process of platinum nanoparticles (Pt NPs) within the Zn-based ZIF-8 Metal-Organic Framework (MOF) \cite{Mechanistic_Poonam} template. The precise understanding and control of this interaction will aid the creation of highly efficient and specialized electrocatalysts that minimize the usage of rare earth elements. 

For generating suggestions for the optimal adsorption positions of Zn ions on the Chini clusters, point generation was guided by the following physical principles: The Zn atom was positioned on the surface of a sphere centered on either Pt or O atoms.
Carbon atoms were excluded as potential adsorption sites since they are unavailable for interaction on the cluster's surface. A mixture of initial simulations and physical intuition determined the sphere's radius. It must be noted that this constraint can err on the high side as the only effect will be an increase in the computational time, with the final energy remaining constant. After evaluating these points, our pipeline suggested the optimal adsorption sites, which were then simulated using DFT. 

The interaction energies at these suggested sites were compared with those obtained from Monte Carlo (MC) sampling. As shown in \Cref{tab:pt-mc-comparison}, the lowest interaction energy predicted by AUGUR is consistently lower than that predicted by the MC approach, a trend that becomes more pronounced with increasing cluster size—from an improvement of 8.73 \% in Pt$_3$ to 142.62 \% in Pt$_9$. This demonstrates that our framework is more robust in finding the global minima than the random MC sampling approach, ultimately saving both time and resources.


        


\begin{table}[ht]
\centering 

\captionsetup{justification=justified}
\caption{Lowest interaction energies for \ch{Pt3}, \ch{Pt6}, and \ch{Pt9} nanoclusters as determined by the Monte Carlo sampling approach and their comparison with the AUGUR pipeline.}
\label{tab:pt-mc-comparison}
\begin{tabularx}{\textwidth}{@{}l *4{>{\centering\arraybackslash}X}@{}}
        & Best Monte Carlo & Best AUGUR & Improvement \\
        & (4 Samples) & (1 Sample \& 2 initial) & [\%] \\
        & {[eV]} & {[eV]} & \\       
\hline
Pt$_{3}$ & -5.32 & -5.79 & 8.73 \\
Pt$_{6}$ & -4.38 & -5.31 & 21.19 \\
Pt$_{9}$ & -1.62 & -3.92 & 142.62 \\
\end{tabularx}
\end{table}

In addition to the optimization process, the high level of symmetry allows for fairly accurate energy predictions. This was evaluated using Leave One Out (LOO) cross-validation. Although the uncertainty was relatively high due to the limited number of points required, the Mean Square Error (MSE), as shown in \cref{tab:pt-loo}, was seen to be remarkably low. A more detailed representation of the LOO can be seen in \Cref{supp-fig:LOO}.
\begin{table}[H]
\caption{Mean Square Error (MSE) for models trained and tested on Pt$_{3}$ and Pt$_{6}$ clusters}
\centering

\begin{tabular}{c|ccc}
                            &        & Tested &        \\
\hline                            
\multirow{3}{*}{\rotatebox[origin=c]{90}{Trained}} & MSE    & \ch{Pt3}    & \ch{Pt6} \\

                            & \ch{Pt3} & 0.26      & 0.66   \\
                            & \ch{Pt6} & 0.05      & 0.07  
\label{tab:pt-loo}
\end{tabular}
\end{table}
%

      
    


Given the accuracy showcased in \Cref{tab:pt-loo}, our model can be leveraged to reconstruct the entire energy surface and quantify uncertainties in unexplored regions. This approach offers a more holistic understanding of the system as compared to the traditional trial and error sampling, which lacks an underlying model for further system interpretation. 

Prior literature knowledge of the electronic structure of Chini clusters allows us to verify the results. These clusters have three distinct chemical sites: the "Top," "Bridged," and "Terminal". Our study shows that the "Top" is the most energetically favorable for Zn ion binding. This is due to the formation of the strong Zn-Pt bonds at the exposed Pt atoms in the outer layer of the cluster, as concluded by our previous study \cite{Mechanistic_Poonam}. Additionally, the "Bridged" and the "Terminal" positions show considerably less strong interaction energies due to the carbonyl ligands inhibiting the direct Metal-Metal (M-M) bonding \cite{Mechanistic_Poonam}. 
AUGUR successfully recreates these findings, accurately identifying the optimal and relative energetics of Zn-chini cluster interactions.

\begin{figure}[H]
    \centering
    \includegraphics[width = \textwidth]{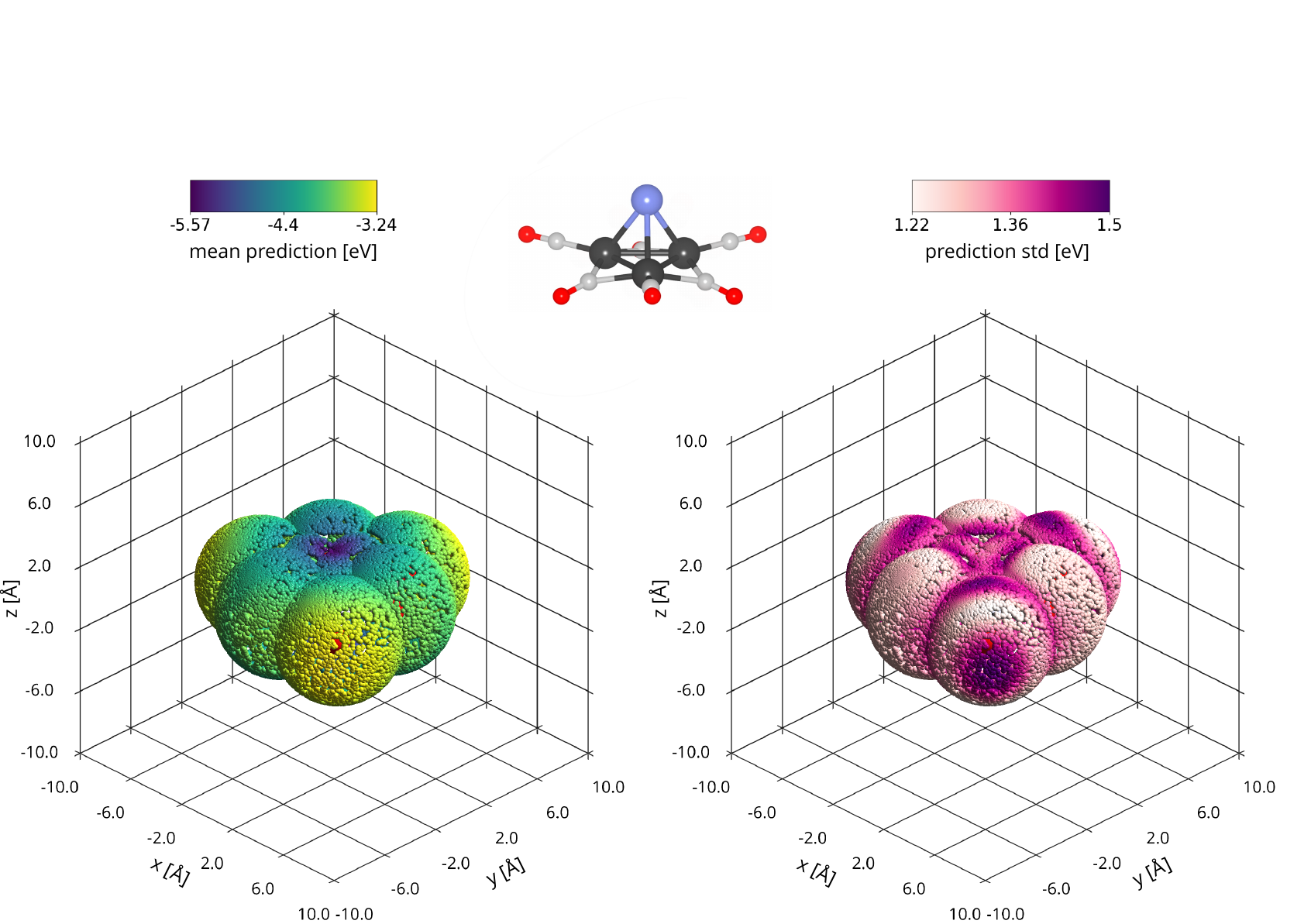}
    \caption{Predicted energy surface for the \ch{Pt3} - Zn cluster in [eV] (left). The cluster figure depicts the most favorable adsorption position as predicted by AUGUR in the Pt$_3$ cluster (middle up). Uncertainty quantification of the prediction (right). The color scheme used is as follows: Pt (black); O (red); C (light grey), Zn (Blue)}
    \label{fig:pt_quality_predictionsPt3}
\end{figure}
In \Cref{fig:pt_quality_predictionsPt3}, the energy surface and optimal site for \ch{Pt3} are presented and verified with the predictions. Of note is that although the predictions are fairly accurate, the uncertainty is relatively high. This can be attributed to the low number of points, as well as to the different distances of Zn ion placement in the training and prediction sets. For a more detailed view of the energy surface and standard deviation, refer to \Cref{supp_fig_pt3_planar}.

\begin{figure}[H]
    \centering
    \includegraphics[width = \textwidth]{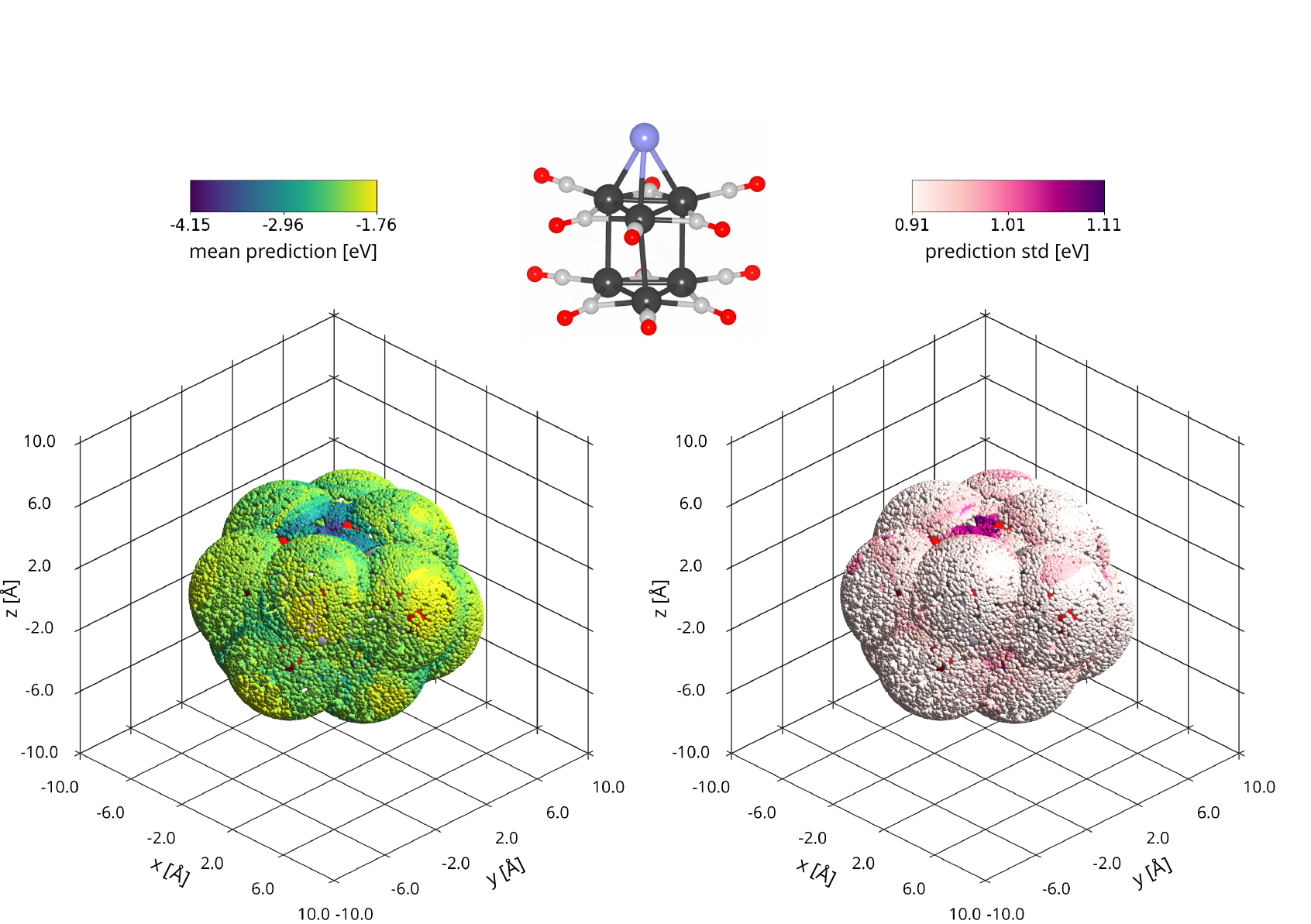}
    \caption{Predicted energy surface for the \ch{Pt6} - Zn cluster in [eV] (left). The cluster figure depicts the most favorable adsorption position as predicted by AUGUR in the Pt$_6$ cluster (middle up). Uncertainty quantification of the prediction (right). The color scheme used is as follows: Pt (black); O (red); C (light grey); Zn (Blue)}
    \label{fig:pt_quality_predictionsPt6}
\end{figure}
In \Cref{fig:pt_quality_predictionsPt6}, the energy surface and optimal site for \ch{Pt6} are presented. The uncertainty is equivalent to that in \Cref{fig:pt_quality_predictionsPt3} despite the larger size. This is because \ch{Pt6} exhibits a strong symmetry, which reduces the number of unique sites to those of \ch{Pt3}. For a more detailed visualization of the energy surface and standard deviation, refer to \Cref{supp_fig_pt6_planar}.
\begin{figure}[H]
    \centering
    \includegraphics[width = \textwidth]{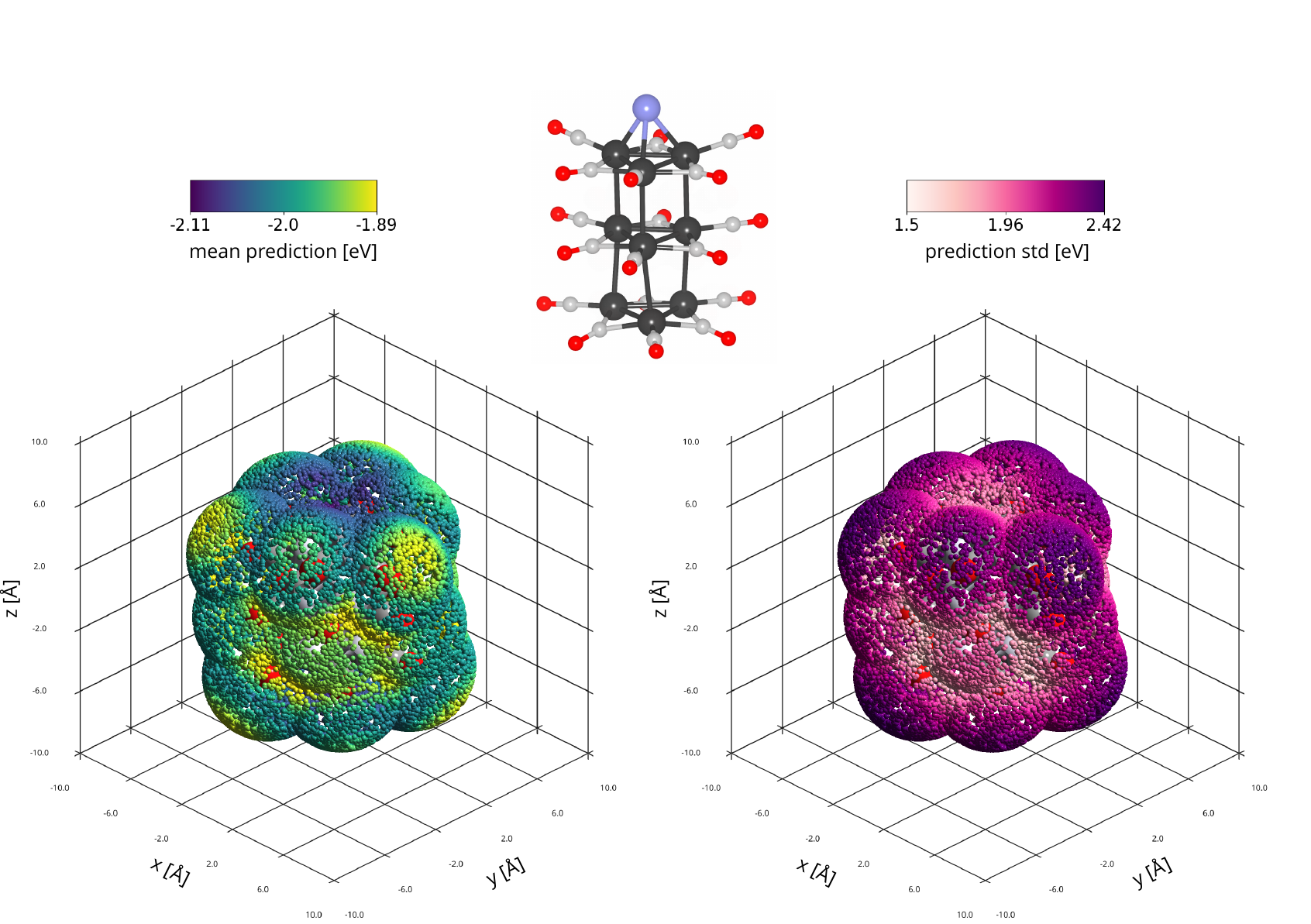}
    \caption{Predicted energy surface for the \ch{Pt9} - Zn cluster in [eV] (left). The cluster figure depicts the most favorable adsorption position as predicted by AUGUR in the Pt$_9$ cluster (middle up). Uncertainty quantification of the prediction (right). The color scheme used is as follows: Pt (black); O (red); C (light grey); Zn (blue)}
    \label{fig:pt_quality_predictionsPt9}
\end{figure}

The \ch{Pt9} cluster introduces two additional distinct sites due to the existence of the middle layer. These are the "Bridged Middle" and "Terminal Middle". According to the prior analyses \cite{Mechanistic_Poonam}, these sites are expected to be less energetically favorable but follow the same energy relationships as those in the upper and lower layers. \Cref{supp_fig_pt9_planar}, provides a detailed view of the energy surface and standard deviation. Of note is that this prediction was generated with only one simulation from \ch{Pt9} included in the training set, alongside \ch{Pt3} and \ch{Pt6}. This finding is of potentially high importance as it suggests that our model could be trained on comparatively inexpensive simulations supplemented by a few strategic points from the more computationally intensive cases to achieve physically consistent and time-efficient results. These models could then be deployed for the prediction and optimization of larger structures of the same family and related systems with higher nuclearity, which would otherwise be computationally infeasible to simulate.   

Nevertheless, it is important to note that the Chini clusters system is relatively simple, and intuition alone would have converged to the optimal site in a reasonable number of iterations. Therefore, our pipeline was subsequently tested on a much more complex system, lacking obvious symmetries.
\FloatBarrier
\subsection{Zinc Oxide Cluster}
\begin{figure}[H]
    \centering
    \includegraphics[scale = 1]{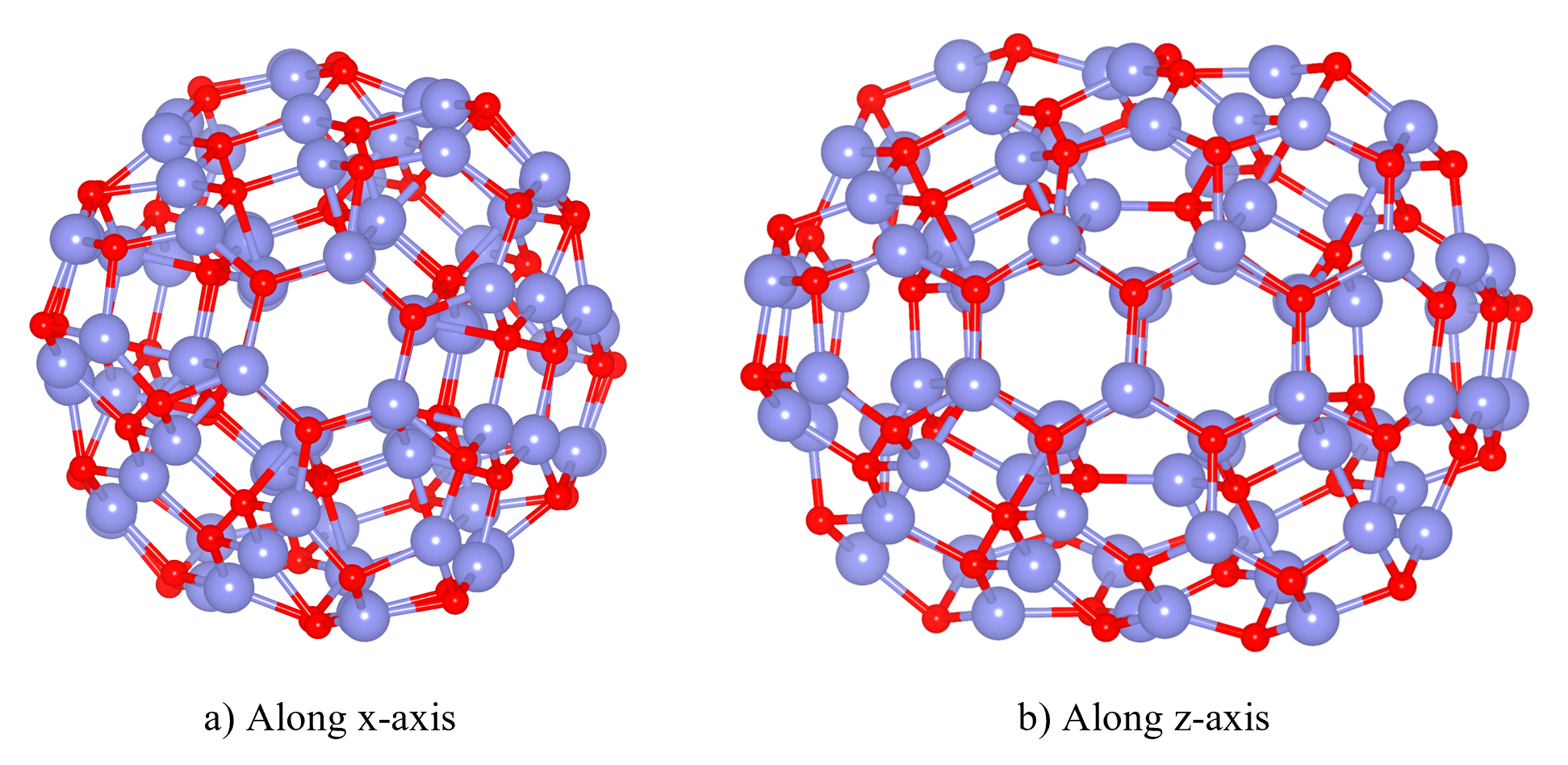}
    \caption{Case Study 2: Zinc Oxide Nanocluster \ch{(ZnO)78}; This figure depicts the relaxed structure of the bare \ch{(ZnO)78} cluster along a) x-axis and b) z-axis. It can be noted that no visible symmetries are seen in the whole cluster. The color scheme used is as follows: O (red); C (light grey); Zn (Blue)} 
    \label{fig:zn_cluster}
\end{figure}

Our second case study involves the Zinc oxide (\ch{(ZnO)}) cluster depicted in \Cref{fig:zn_cluster}, with carbon monoxide (CO) as the adsorbate molecule. ZnO clusters are used experimentally and industrially in semiconductor companies for chips, sensors, and electrodes due to their observed high catalytic activity and stability \cite{Zno_applications}. Our study aims to aid the exploration of the gas adsorption potential of the large Zn oxide clusters, which would otherwise be very computationally expensive to pursue. The \ch{(ZnO)78} was mainly selected because it is one of the "magic clusters"\cite{MagicZnO78}, yet remains relatively unexplored in the literature. Given the complexity and asymmetry of such an extensive system, manually identifying all the chemically distinct adsorption sites on the potential energy surface (PES) to begin the exploration is impractical. Through our investigation, we discovered two distinct sites in what appeared to be nearly identical locations. However, the adsorption energies varied significantly between these sites, with one being the optimal adsorption position and the other showing only half the absolute adsorption energy. 

The point generation in this study was done by positioning the carbon atom of the adsorbate on the surface of a sphere centered on a randomly selected atom within the cluster, with a radius of \SI{1.4}{\angstrom}. Subsequently, the oxygen atom of the adsorbate was placed on the surface of a sphere centered on the carbon atom, with a radius of \SI{1.12}{\angstrom}, corresponding to a triple bond between the C and O atoms of carbon monoxide. The placement was further constrained to maintain a minimum distance of \SI{1.4}{\angstrom} between the oxygen atom and the Zinc oxide cluster so that it does not coincide with the cluster atoms. Additionally, the carbon atom was positioned first as it shows a higher chemical affinity for the cluster. In this way, we ensure that our method does not preclude placements of the adsorbate where the O atom is closer to the cluster but makes it less likely.  



We present the comparison of energy optimization results obtained using both Monte Carlo (MC) sampling and the AUGUR framework in \Cref{tab:zn-mc-comparison}. In total, 19 samples were evaluated with the MC approach, compared to 13 with AUGUR. Despite the lower number of samples, the AUGUR framework consistently outperformed the MC method in both energy optimization and data efficiency. To ensure that the optimum identified by AUGUR is indeed the global minimum, approximately 70 additional simulations were conducted. After thoroughly investigating the structure, we are confident that the identified optimum is the global minimum.


\begin{table}[ht]
\centering 
\captionsetup{justification=justified}
\caption{Lowest interaction energies and Mean Square Error (MSE) results of the \ch{(ZnO)78} clusters as determined by the Monte Carlo (MC) sampling approach and their comparison with the AUGUR pipeline.} 
\label{tab:zn-mc-comparison}
\begin{tabularx}{\textwidth}{@{}l *4{>{\centering\arraybackslash}X}@{}}
        & Monte Carlo & AUGUR & Improvement \\
        & (19 samples) & (13 samples ) & [\%] \\
        & {[eV]} & {[eV]} & \\           
\hline
Lowest energy& -0.810      & -0.902& 11.3 \\
Model cross-validation MSE   &1.77 &0.63 &64.4 \\
\end{tabularx}
\end{table}

\begin{figure}[H]
    \subcaptionbox*{cross validation error using the MC samples as dataset)}[.50\linewidth]{%
    \includegraphics[width=\linewidth]{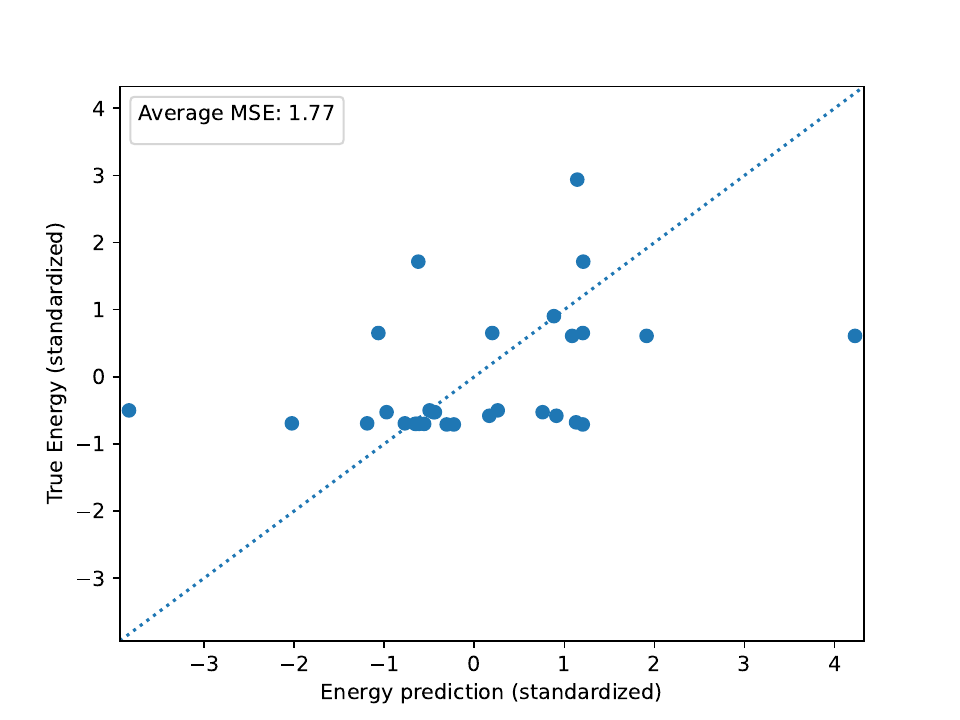}%
  }
  \hfill
    \subcaptionbox*{cross validation error using the AUGUR samples as dataset}[.50\linewidth]{%
    \includegraphics[width=\linewidth]{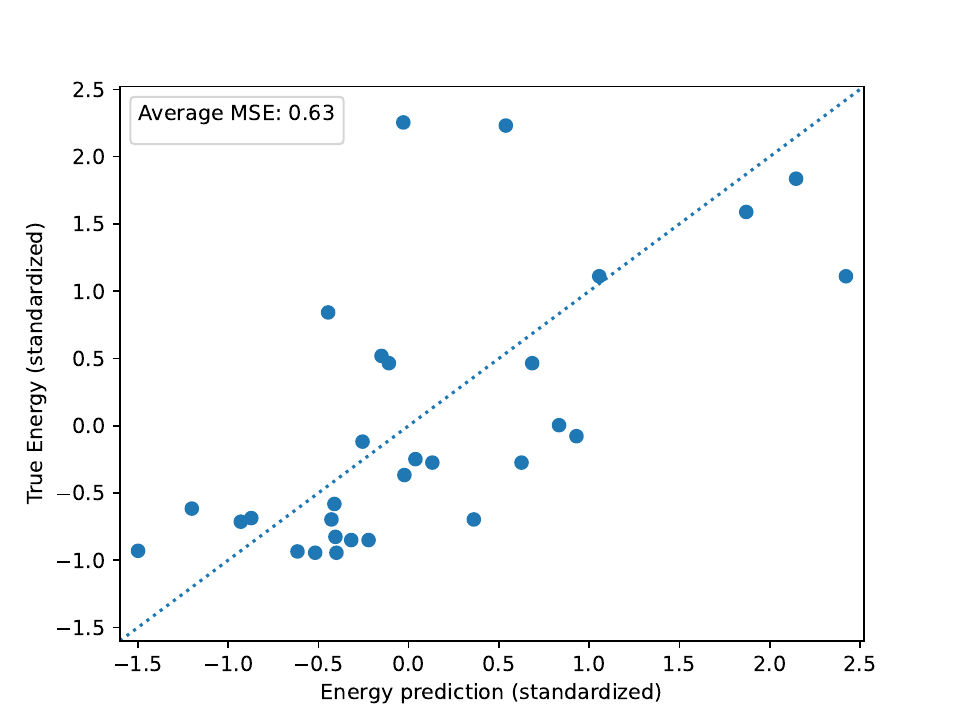}%
  }%
    \hfill
  \caption{Quality of best site and model training using MC collected samples.}
    \label{fig:augur_comparison_mse}
\end{figure}

An additional advantage of this approach is its ability to generate data points that train an accurate model in the most data-efficient manner. To demonstrate this, we present the cross-validation results from both approaches in \Cref{fig:augur_comparison_mse}. Each AUGUR step involved two suggestions: one focused on extreme exploration (with a tradeoff of 100) and the other on extreme exploitation (with a tradeoff of 0.5). This method could be further refined by gradually decreasing the tradeoff and generating only one suggestion per step. However, the combined approach enabled rapid convergence—not in terms of the total number of simulations but in terms of the number of suggestion steps—by allowing the two extremes to be evaluated in parallel.

\begin{figure}[H]
    \centering
    \includegraphics[width = \textwidth]{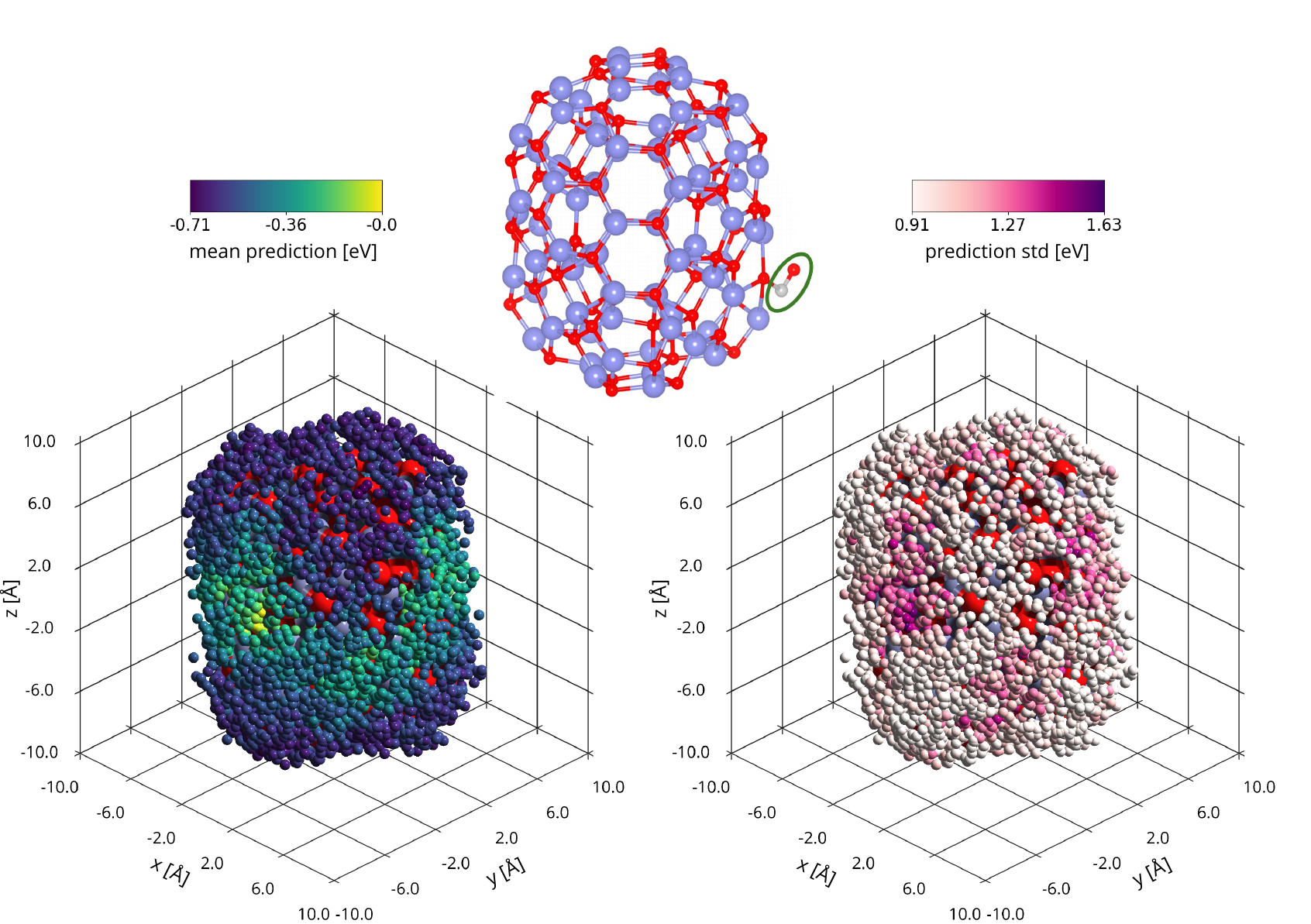}
      \caption{Results for the energy surface of the \ch{(ZnO)78} cluster (left), the most favorable adsorption position (middle up) and the corresponding uncertainty (right).}
    \label{fig:full_investigation_zn}
\end{figure}

In \Cref{fig:full_investigation_zn}, we show the predicted energy surfaces and associated uncertainties of the investigated cluster by predictions made at 5000 points. The prediction of those suggested points required approximately 150 seconds. In addition, the average simulation duration for one suggestion of this case study is approximately two days. Therefore, performing an exhaustive energy surface investigation using traditional means would have taken multiple years. 



For a more detailed view of the energy surface and standard deviation, refer to \Cref{supp_fig_zno_planar}. 
This case study, more than anything, highlights the advantages of the AUGUR framework. Our pipeline not only determines the best energy site with minimal data cost, it also gathers points efficiently for the model's training which can be seamlessly used to gain physical insight that would have otherwise taken years to achieve.

\FloatBarrier
\section{Outlook}
In summary, we have presented a novel deep kernel learning framework for automatically determining the optimal adsorption sites on molecules, minimizing the number of DFT simulations required. It seamlessly combines graph neural networks and Gaussian processes. In this way, we achieve rotation, symmetry, and translation invariance for the inputs of the Gaussian processes. Additionally, because of the properties of graphs, we also allow for the same model to be used across different molecules regardless of their size. This alleviates the need for crude techniques such as padding or hand-crafted features that require a high level of physical intuition. Further, this allows for significant dimensionality reduction to the inputs of the Gaussian processes, which famously scale poorly with increasing feature dimensions. This composite model will enable us to fully investigate the energetic structure of the molecule as well as selectively refine areas of high uncertainty if deemed necessary. 
The model is then used as the surrogate to a Bayesian Optimization scheme that iteratively refined both the model and the suggested sites for optimal adsorption till it converged to the global optimum. 

It is shown that this algorithm performs robustly and efficiently, determining both the optimal sites as well as predicting the energy surface with a limited number of datapoints. 
The natural next step is to deploy this algorithm for cases of highly complex adsorbents and leverage AUGUR to further our chemical understanding of hitherto computationally prohibitive systems.

\printbibliography
\section*{ACKNOWLEDGEMENTS}
  The authors acknowledge funding from the project ProperPhotoMile, supported under the umbrella of SOLAR-ERA.NET Cofund 2 by The Spanish Ministry of Science and Education and the AEI under the project PCI2020-112185 and CDTI project number IDI-20210171; the Federal Ministry for Economic Affairs and Energy on the basis of a decision by the German Bundestag project number FKZ 03EE1070B and FKZ 03EE1070A; and the Israel Ministry of Energy with project number 220-11-031. SOLAR-ERA.NET is supported by the European Commission within the EU Framework Programme for Research and Innovation HORIZON 2020 (Cofund ERA-NET Action, N° 786483).\\
  Further, A.G. acknowledges financial support from TUM Innovation Network for Artificial Intelligence powered Multifunctional Material Design (ARTEMIS) and funding in the framework of Deutsche Forschungsgemeinschaft (DFG, German Research Foundation) under Germany's Excellence Strategy – EXC 2089/1 – 390776260 (e-conversion).
  Lastly, we wish to express our gratitude to Dr. Inigo Iribarren for creating the flowcharts for this work.

\end{document}